\documentclass[prl,reprint,twocolumn,showpacs,superscriptaddress]{revtex4-2}
\usepackage[export]{adjustbox}
\usepackage{amsmath,amssymb,bm,mathrsfs,graphicx, braket, times, mathtools}
\usepackage[colorlinks=true,citecolor=blue,linkcolor=blue]{hyperref}
\usepackage{multirow}
\usepackage[all,cmtip]{xy}
\usepackage[normalem]{ulem}
\usepackage[usenames,dvipsnames]{color}
\usepackage{bbding}
\usepackage{graphicx}
\usepackage{subfigure}

\usepackage[utf8]{inputenc}
\usepackage[english]{babel}
\usepackage{amsthm}
\usepackage{wasysym}

\begin{document}

\title{Hidden Quantum Geometry in Bilayer Exciton Condensates}

\author{Xuzhe Ying}
\affiliation{Department of Physics, Hong Kong University of Science and Technology, Clear Water Bay, Hong Kong, China}
\affiliation{Center for Theoretical Condensed Matter Physics, The Hong Kong University of Science and Technology, Clear Water Bay, Hong Kong, China}
\author{Benjamin T. Zhou} \thanks{Contact author: tongz@hkust-gz.edu.cn}
\affiliation{Quantum Science and Technology Center \& Thrust of Advanced Materials,
The Hong Kong University of Science and Technology (Guangzhou), 1 Duxue Rd., Nansha, Guangzhou, China}

\begin{abstract}
When an electron-doped layer is stacked with a hole-doped layer with approximately equal carrier density, inter-layer Coulomb interaction turns the bilayer system into an exciton condensate (EC). In this Letter, we reveal a fundamental property of bilayer ECs: the excitonic order gives rise to nontrivial hidden quantum geometric effects in the correlated electron-hole bands, even when the non-interacting bands are trivial. Such peculiar EC-driven quantum geometry manifests itself in a characteristic \emph{out-of-plane} polarization response upon applying an \emph{in-plane} AC electric field to the bilayer EC system. In particular, the second-order response exhibits a characteristic inverse square scaling with the bilayer EC order parameter. Our finding reveals a fundamental hidden Berry phase effect driven by electron-hole correlations, and establishes bilayer EC as a promising platform for rich nonlinear physics.
\end{abstract}

\maketitle

\emph{Introduction---}As a prototypical correlated phase of matter, exciton condensates (ECs) have been one of the central topics of condensed matter physics over the past few decades \cite{Eisenstein2014ExcitonCondensate,Eisenstein2004BECExciton,PhysRevLett.68.1383,PhysRevLett.68.1379}, which exhibits peculiar transport phenomena such as perfect Coulomb and Hall drag~\cite{PhysRevLett.88.126804,PhysRevLett.90.246801,Nandi2012PerfectCoulombDrag} and dissipationless counterflow~\cite{MACDONALD2001129,PhysRevLett.93.036801,PhysRevLett.93.036802}.
Recent experimental discoveries of bilayer ECs in a rich variety of graphene and transition-metal dichalcogenide bilayer systems~\cite{Wang2019EvidenceEC,Ma2019StrongEI,Gu2022Dipolar,PhysRevLett.126.106804,Shi2022BilayerWSe2,zeng2026Observation,wang2018Electrical,doi:10.1126/science.aaw4194,doi:10.1126/science.abg1110} have triggered extensive search for exotic inter-layer excitonic states \cite{PhysRevLett.121.126601,PhysRevB.105.235121,PhysRevLett.126.137601,xs7g-dmn8,PhysRevX.13.031023,doi:10.1073/pnas.2424663122}. 

Recent studies have seen a surge of interest in quantum geometric effects in bilayer ECs \cite{PhysRevLett.127.170404,PhysRevLett.132.236001,PhysRevB.105.L140506,PhysRevB.111.014502,ying2024flatbandexcitonsquantum,3pvf-1ckd}. However, Refs.~\cite{PhysRevLett.127.170404,PhysRevLett.132.236001,PhysRevB.105.L140506,PhysRevB.111.014502,ying2024flatbandexcitonsquantum,3pvf-1ckd} focus on the quantum geometry readily encoded in non-interacting electron and hole states. Indeed, in multi-orbital lattice models (such as Lieb lattice or twisted bilayer graphene), the non-interacting electronic wavefunctions already possess nontrivial quantum geometry. It has been shown that the phase stiffness \cite{PhysRevLett.132.236001,3pvf-1ckd} and the stability \cite{PhysRevLett.127.170404,PhysRevB.105.L140506,3pvf-1ckd} of bilayer ECs constructed out of the flat-band models are highly sensitive to the non-interacting electronic quantum metric.

Here we investigate a more fundamental aspect regarding quantum geometry in bilayer ECs: whether excitonic order itself generates intrinsic quantum geometry in a bilayer EC even when the non-interacting electron and hole bands are trivial. If yes, what are the physical consequences of the EC induced quantum geometry? This is motivated by the hallmark of a bilayer EC, namely the spontaneous inter-layer coherence driven by Coulomb interactions~\cite{Eisenstein2014ExcitonCondensate,Eisenstein2004BECExciton,PhysRevLett.84.5808}. In the mean field description, the order parameter \emph{coherently} hybridizes electron and hole states across the two layers, reconstructing the quasi-particle wavefunctions and thereby inducing potential quantum-geometric effects absent in the bare bands.


In this Letter, we substantiate the observation above: the inter-layer EC order generates \emph{hidden} quantum geometry with significant physical consequences in the simplest bilayer EC, where both the electron and hole bands are utterly trivial. In addition to the usual Berry curvature \cite{RevModPhys.82.1959,vanderbilt2018berry} and quantum metric \cite{provost1980riemannian,cheng2010quantum}, one can define the \emph{in-plane Berry curvature} of the bilayer EC, which is ``hidden'' due to the lack of translation symmetry in the out-of-plane direction. The generalized quantum geometric tensors necessarily involves matrix elements of the $z$-position operator, and they are indexed by three spatial directions. Using the quantum kinetic equation framework \cite{x9pk-lb4x,PhysRevB.109.235137,kamenev2023field}, we reveal that the hidden quantum geometry in bilayer ECs manifests itself in the response function of the \emph{out-of-plane} polarization $P_z$ to an applied \emph{in-plane} AC electric field. Specifically, an in-plane AC field drives an out-of-plane dipole oscillation at second-harmonic frequencies. 

Importantly, for the ground state of bilayer EC, the second-order dipole response function due to EC-driven quantum geometry has a peculiar inverse-square power law dependence on the EC order parameter: $\chi^{(2)}_{z;\alpha\beta} \propto \Delta^{-2}$. Thus, the correlation-driven response cannot be found via a perturbative expansion in terms of the EC order parameter. This unique non-perturbative nature can serve as a distinctive experimental signature for the inter-layer coherence and EC order parameter.

\begin{figure}[t]
    \centering
    \includegraphics[width=\linewidth]{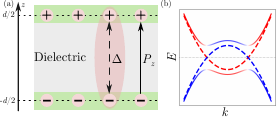}
    \caption{Bilayer system and exciton condensate. (a) Schematic system setup with some electrons (holes) in the bottom (top) layer. The out-of-plane polarization $P_z$ will exhibit a distintive response to the in-plane AC field. (b) Schematic quasi-particle spectrum (solid line). Dashed lines correspond to the non-interacting electronic bands. The mean field order parameter $\Delta$ hybridizes the electrons' wavefunctions in the two layers.}
    \label{fig:BilayerSystem}
\end{figure}


\emph{Bilayer exciton condensate and hidden Berry curvature---}In this work, we take the bilayer electronic system in Fig.~\ref{fig:BilayerSystem}(a) as a representative example of two dimensional stacked materials. In Fig.~\ref{fig:BilayerSystem}(a), there are some electrons (holes) in the bottom (top) layer. Due to the Coulomb interaction, the electron and hole form the boundstate of exciton, which condenses into a superfluid state at low temperature.

We focus on the low-temperature phase of exciton condensation, which spontaneously breaks the relative $U(1)$ symmetry. Namely, the charge for each layer is no longer conserved, while total charge remains conserved. In the mean field description, the Hamiltonian is given  by:
\begin{equation}
    \hat{H}_0(\boldsymbol{k})=\begin{bmatrix}
        c^{\dagger}_{\boldsymbol{k}} & d^{\dagger}_{\boldsymbol{k}}
    \end{bmatrix}\begin{bmatrix}
        -\xi(\boldsymbol{k})+\mu & \Delta\\
        \Delta & \xi(\boldsymbol{k})-\mu
    \end{bmatrix}\begin{bmatrix}
        c_{\boldsymbol{k}}\\
        d_{\boldsymbol{k}}
    \end{bmatrix}
    \label{Eq:EC_MF}
\end{equation}
where $c_{\boldsymbol{k}}$ and $d_{\boldsymbol{k}}$ are the electron annihilation operators at momentum $\boldsymbol{k}$ for top and bottom layers separately. The mean field order parameter $\Delta$ coherently couples the electrons in the two layers. The quasi-particle's spectra is given by $\epsilon_{n}(\boldsymbol{k})$, with a gap shown in Fig.~\ref{fig:BilayerSystem}(b). In addition, the quasi-particle's wavefunction is now a hybridization of electronic wavefunction from both layers.

The out-of-plane polarization (OPP) in Fig.~\ref{fig:BilayerSystem}(a) is defined as $P_z:=-\frac{1}{2}\frac{1}{V}\sum_{\boldsymbol{r}}\left[\langle c^{\dagger}_{\boldsymbol{r}}c_{\boldsymbol{r}}\rangle-\langle d^{\dagger}_{\boldsymbol{r}}d_{\boldsymbol{r}}\rangle\right]$, where $V$ is the system size and the unit of $ed$ (electronic charge times layer separation) is suppressed. In the bilayer system, the out-of-plane polarization is essentially the difference in the electron density of the two layers. Given the self-consistent mean field equations, the out-of-plane polarization can be further computed by the quasi-particle wavefunctions:
\begin{equation}
    P_z=\sum_{n=\pm}\int \frac{d^2k}{(2\pi)^2}u^{\dagger}_{n}(\boldsymbol{k})\hat{z}\ u_n(\boldsymbol{k})f_n^{\text{eq}}(\boldsymbol{k})
    \label{Eq:Out_of_plane_Polarization}
\end{equation}
where the summation $n=\pm$ is over the quasi-particle bands; $u_n(k)$ is the periodic part of the Bloch wavefunction in band ``$n$'' and momentum $\boldsymbol{k}$ and $f_n^{\text{eq}}(k)$ is the corresponding equilibrium Fermi distribution function. Here, the position operator $\hat{z}$ in the out-of-plane direction is given by the Pauli matrix $\sigma_z$ in the layer basis, namely $\hat{z}=-\frac{1}{2}\sigma_z$.

In the modern theory of polarization \cite{resta2007theory,PhysRevB.47.1651,RevModPhys.66.899,vanderbilt2018berry}, the \emph{in-plane} polarization is closely related to the Berry phase or integration of Berry connection over Brillouin zone. With such analogy, we may think of the \emph{out-of-plane} ``Berry connection'' as:
\begin{equation}
    A^z_n(\boldsymbol{k})=u^{\dagger}_n(\boldsymbol{k})\ \hat{z}\ u_n(\boldsymbol{k})
    \label{Eq:OutofPlane_BerryConnection}
\end{equation}
where the integration of $A^z_n(\boldsymbol{k})$ gives the out-of-plane polarization $P_z$ in Eq.~(\ref{Eq:Out_of_plane_Polarization}). Besides the superficial analogy, the out-of-plane ``Berry connection'' and the \emph{in-plane} ``Berry curvature'' (to be defined below) can be put on equal physical footing with the usual Berry connection and Berry curvature. This equivalence has been established in recent studies of bilayer ferroelectrics~\cite{PhysRevLett.132.196801,https://doi.org/10.1002/adfm.202408625}.

In the Bloch state basis, the position operator projected onto $n$-th band is given by $\hat{r}_{\alpha,n}=i\partial_{k_{\alpha}}+A^{\alpha}_n(\boldsymbol{k})$ for in-plane directions $\alpha=x,y$ and $\hat{r}_{z,n}=A_n^z(\boldsymbol{k})$ for the out-of-plane $z$-direction. The velocity operator follows from the Heisenberg equation $\hat{v}_{\alpha,n}=-i\frac{d}{dt}\hat{r}_{\alpha,n}=-i\left[\hat{r}_{\alpha,n}, \epsilon_n(\boldsymbol{k})+eE_{\beta}\ \hat{r}_{\beta,n}\right]$. In the in-plane direction, this equation yields the standard semi-classical motion of Bloch electrons with group velocity $v_{\alpha,n}^{\text{g}}=\partial_{k_{\alpha}}\epsilon_n(\boldsymbol{k})$ and the anomalous velocity $v_{\alpha,n}^{\text{a}}=e\epsilon_{\alpha\beta\gamma}E_{\beta}\ \Omega^n_{\gamma}(\boldsymbol{k})$ from Berry curvature \cite{RevModPhys.82.1959,vanderbilt2018berry}.

In the out-of-plane $z$-direction, the group velocity vanishes and only the anomalous part survives $v_{z,n}=e\epsilon_{z\beta\gamma}E_{\beta}\ \Omega^n_{\gamma}(\boldsymbol{k})$. This results in the \emph{hidden} in-plane component of Berry curvature, defined as 
\begin{equation}
    \Omega^n_{\alpha}(\boldsymbol{k}):=\epsilon_{\alpha\beta z}\partial_{k_{\beta}}A^z_n(\boldsymbol{k}).
\end{equation}
for $\alpha,\beta=x,y$. The velocity in $z$-direction describes the inter-layer charge transfer. To see this point, we utilize the kinetic equation $\partial_t f_n-e\boldsymbol{E}\cdot\partial_{\boldsymbol{k}}f_n=0$ for spatially uniform distribution function $f_n$ in the \emph{absence} of scattering. Then, it is straightforward to compute the changing rate of out-of-plane polarization $\frac{d}{dt}P_z(t)=\sum_n\int\frac{d^2k}{(2\pi)^2}A_n^z(\boldsymbol{k})\partial_tf_n$. The result is given by $\frac{d}{dt}P_z(t)=\sum_n\int\frac{d^2k}{(2\pi)^2}e\epsilon_{z\beta\gamma}E_{\beta}\ \Omega^n_{\gamma}(\boldsymbol{k})f_n$, which is a purely kinematic effect.

In the presence of scattering, the story is slightly different. Imagine there is certain scattering mechanism, described by a relaxation time $\tau$. Then, under the external driving field $\boldsymbol{E}$, the quasi-particle's distribution is approximately given by $f_n(\boldsymbol{k})\approx f_n^{\text{eq}}(\boldsymbol{k})+e\tau\boldsymbol{E}\cdot\partial_{\boldsymbol{k}}f_n^{\text{eq}}(\boldsymbol{k})$ in the limit of low driving frequency. It is straightforward to compute the out-of-plane polarization. The change in the out-of-plane polarization due to external field is given by
\begin{equation}
    \delta P_z=-e\tau\sum_{n}\int\frac{d^2k}{(2\pi)^2}\boldsymbol{\hat{e}}_z\cdot\left(\boldsymbol{E}\times\boldsymbol{\Omega}^n\right)f_n^{\text{eq}}(\boldsymbol{k})
    \label{Eq:ChangeOPP_LinearResponse}
\end{equation}
where $\boldsymbol{\hat{e}}_z$ is the unit vector in $z$-direction. In the bilayer system, instead of the current, the hidden in-plane Berry curvature shows up in the change of out-of-plane polarization under external bias.

Note that in general, the change in the polarization in Eq.~(\ref{Eq:ChangeOPP_LinearResponse}) is not the full story. The electric (scalar) potential generically breaks the translation symmetry. Hence, the bare Bloch wavefunction is no longer a good basis in a technical sense. Below, we systematically find the modifications to the wavefunctions due to the electric field, which further contributes to the response function. In computing the response functions, additional generalized quantum geometric tensors show up. Intriguingly, the second order response has a nontrivial characterisitic dependence on the EC order parameter $\Delta$, providing a direct probe to the inter-layer coherence.

\emph{Nonlinear response to in-plane field---}We utilize the kinetic equation \cite{x9pk-lb4x,PhysRevB.109.235137,kamenev2023field} to develop a systematic response theory of the out-of-plane polarization $P_z(t)$ up to second order in the in-plane AC field $\boldsymbol{E}(t)$. For concreteness, we consider the bilayer system biased by a uniform driving field with the following real-space Hamiltonian:
\begin{equation}
    \hat{H}(\boldsymbol{r},\boldsymbol{r}^{\prime};t)=H_0(\boldsymbol{r}-\boldsymbol{r}^{\prime})+e\boldsymbol{E}(t)\cdot\boldsymbol{r}\delta_{\boldsymbol{r},\boldsymbol{r}^{\prime}}
\end{equation}
where $H_0(\boldsymbol{r}-\boldsymbol{r}^{\prime})$ describes the quasi-particle's kinetic energy on a lattice. One can systematically access the semi-classical limit by Wigner transformation \cite{kamenev2023field,x9pk-lb4x}, i.e., the Fourier transformation on the relative coordinate $(\boldsymbol{r}-\boldsymbol{r}^{\prime})$. After Wigner transformation, we obtain the Hamiltonian in \emph{phase} space, parameterized by position $\boldsymbol{r}$, momentum $\boldsymbol{k}$, and time $t$:
\begin{equation}
    \hat{H}(\boldsymbol{k},\boldsymbol{r};t)=H_0(\boldsymbol{k})+e\boldsymbol{E}(t)\cdot\boldsymbol{r}
    \label{Eq:Ham_PhaseSpace}
\end{equation}
To define out-of-plane polarization (or any other observables), we need to deduce two additional pieces of information from the Hamiltonian in Eq.~(\ref{Eq:Ham_PhaseSpace}). Namely, we will need the wavefunctions $\mathbb{U}_n(\boldsymbol{k},\boldsymbol{r};t)$ and the kinetic equation for the distribution function $f_n(\boldsymbol{k},\boldsymbol{r};t)$. Significant simplification can be made by considering the uniform electric field as assumed above. The assumption of uniform electric field holds as long as it varies slowly on the scale of correlation length or mean free path.

Under a uniform electric field, we consider the spatially uniform distribution function  $f_n(\boldsymbol{k},\boldsymbol{r};t)=f_n(\boldsymbol{k};t)$, which satisfies the usual Boltzmann equation:
\begin{equation}
    \partial_tf_n(\boldsymbol{k};t)-e\boldsymbol{E}(t)\cdot\partial_{\boldsymbol{k}}f_n(\boldsymbol{k};t)=-\frac{f_n(\boldsymbol{k};t)-f_n^{\text{eq}}(\boldsymbol{k})}{\tau}
\end{equation}
where the right hand side is the collision integral within relaxation time approximation. For exciton insulators, we can just take the distribution function to be the equilibrium Fermi distribution $f_n^{\text{eq}}(\boldsymbol{k})$. This is fine in the low driving frequency limit without resonance and at low temperature where the distribution function $f_n(\boldsymbol{k})$ is almost constant.

Complication comes from the wavefunction $\mathbb{U}_n(\boldsymbol{k},\boldsymbol{r};t)=\mathbb{U}_n(\boldsymbol{k};t)$, which can be constructed to be position independent in the current situation (see Supplemental Materials (SM) \cite{SupplementalM}). Note that the electric potential explicitly breaks the translation symmetry of the lattice. Hence, the bare Bloch wavefunction is no longer a good basis. Instead, we look for eigen-basis defined by the following Moyal equation:
\begin{equation}
    \hat{H}(\boldsymbol{k},\boldsymbol{r};t)\star\mathbb{U}_n(\boldsymbol{k},\boldsymbol{r};t)=\mathbb{U}_n(\boldsymbol{k},\boldsymbol{r};t)\star\tilde{h}_n(\boldsymbol{k},\boldsymbol{r};t)
\end{equation}
with the Moyal star operation defined as $\star=\text{Exp}[-\frac{i}{2}(\overset{\leftarrow}{\partial_{\boldsymbol{r}}}\cdot\overset{\rightarrow}{\partial_{\boldsymbol{k}}}-\overset{\leftarrow}{\partial_{\boldsymbol{k}}}\cdot\overset{\rightarrow}{\partial_{\boldsymbol{r}}})]$, encoding the non-commutative phase-space structure. The solution can be found perturbatively in the space and momentum derivatives (under the name of gradient expansion). The solution up to second order in electric field $\boldsymbol{E}$ is provided in the SM \cite{SupplementalM}.

Lastly, given the distribution function $f_n(\boldsymbol{k};t)$ and the wavefunction $\mathbb{U}_n(\boldsymbol{k};t)$, observables can be constructed. In the current case, the out-of-plane polarization is given by:
\begin{equation}
    P_z(t)=\sum_{n}\int\frac{d^2k}{(2\pi)^2}\mathbb{U}_n^{\dagger}(\boldsymbol{k};t)\ \hat{z}\ \mathbb{U}_n(\boldsymbol{k};t)\ f_n(\boldsymbol{k};t).
    \label{Eq:Out-of-plane_Polarization_KE}
\end{equation}
Generally, the observables are not necessarily given by this simple formula. One has to carefully construct gauge-invariant quantities, see Ref.~\cite{x9pk-lb4x}. Eq.~(\ref{Eq:Out-of-plane_Polarization_KE}) is the correct result for the current case. Below, we focus on the exciton insulators and take $f_n(\boldsymbol{k};t)=f_n^{\text{eq}}(\boldsymbol{k})$.

\begin{figure}[h]
    \centering
    \includegraphics[width=\linewidth]{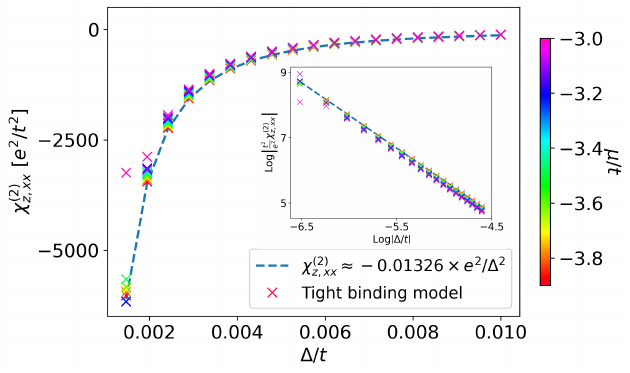}
    \caption{Exciton condensate's nonlinear out-of-plane dielectric response $\chi^{(2)}_{z,xx}$ v.s. mean field order parameter $\Delta$ at various ``chemical potentials'' $\mu$. Inset: The same data (taking absolute value) plotted on log-log scale.}
    \label{fig:EC_chi}
\end{figure}

\begin{widetext}
Linear response of out-of-plane polarization to the in-plane field is given by:
\begin{equation}
    P_{z}^{(1)}(t)=e E_{\alpha}(t)\sum_{n}\sum_{m\neq n}
    \int\frac{d^2k}{(2\pi)^2}\frac{\hat{z}_{nm}(\boldsymbol{k})A^{\alpha}_{mn}(\boldsymbol{k})+A^{\alpha}_{nm}(\boldsymbol{k})\hat{z}_{mn}(\boldsymbol{k})}{\epsilon_m(\boldsymbol{k})-\epsilon_n(\boldsymbol{k})} f_n^{\text{eq}}(\boldsymbol{k})
    \label{Eq:LinearResponse}
\end{equation}
where $\hat{z}_{mn}(\boldsymbol{k})=u_m^{\dagger}(\boldsymbol{k})\hat{z}\ u_n(\boldsymbol{k})$ and $A^{\alpha}_{mn}=u_m^{\dagger}(\boldsymbol{k})i\partial_{k_{\alpha}}u_n(\boldsymbol{k})$. The integrand in Eq.~(\ref{Eq:LinearResponse}) can be viewed as generalized quantum metric. This linear response result matches the Kubo formula result when the driving frequency is much lower than the gap, see SM \cite{SupplementalM}. In the simple example to be considered below, the linear response vanishes due to rotation symmetry. Hence, below we proceed to derive second order response function. Generically, linear response function should be nonzero in systems with low symmetry, for instance those without rotation and time reversal symmetries.

Higher order response is generally more complicated, involving out-of-time-order correlations \cite{PhysRevB.101.045201}. Indeed, the higher order response function generically involves nonequilibrium processes, due to the lack of fluctuation-dissipation theorem beyond linear response \cite{altland2010condensed}. Nonetheless, the kinetic equation \cite{x9pk-lb4x,PhysRevB.109.235137,kamenev2023field}, as deducible from nonequilibrium field theory \cite{kamenev2023field}, necessarily captures the nonequilibrium processes and is ideal for computing higher order response functions.

Indeed, the second order response can be straightforwardly computed from Eq.~(\ref{Eq:Out-of-plane_Polarization_KE}) by truncating it at second order in electric field, namely $P_z^{(2)}(t)=\chi_{z;\alpha\beta}^{(2)}E_{\alpha}(t)E_{\beta}(t)$. The response function $\chi_{z;\alpha\beta}^{(2)}$ is now given by
\begin{equation}
\resizebox{\linewidth}{!}{
$\begin{split}
    \chi_{z;\alpha\beta}^{(2)}/e^2=&\frac{1}{2}\sum_{n}\int\frac{d^2k}{(2\pi)^2}f_n^{\text{eq}}(\boldsymbol{k})\sum_{m_1,m_2\neq n}\frac{A^{\alpha}_{nm_1}(\boldsymbol{k})\hat{z}_{m_1m_2}(\boldsymbol{k})A^{\beta}_{m_2n}(\boldsymbol{k})}{\left[\epsilon_{m_1}(\boldsymbol{k})-\epsilon_{n}(\boldsymbol{k})\right]\left[\epsilon_{m_2}(\boldsymbol{k})-\epsilon_{n}(\boldsymbol{k})\right]}
    -\frac{1}{2}\sum_n\int\frac{d^2k}{(2\pi)^2}f_n^{\text{eq}}(\boldsymbol{k})\hat{z}_{nn}(\boldsymbol{k})\sum_{m\neq n}\frac{A^{\alpha}_{nm}(\boldsymbol{k})A^{\beta}_{mn}(\boldsymbol{k})}{\left[\epsilon_{m}(\boldsymbol{k})-\epsilon_n(\boldsymbol{k})\right]^2}\\
    +&\frac{1}{2}\sum_{n}\int\frac{d^2k}{(2\pi)^2}f_n^{\text{eq}}(\boldsymbol{k})\sum_{m\neq n}\hat{z}_{nm}\left\{\sum_{l\neq n}\frac{A^{\alpha}_{ml}(\boldsymbol{k})A^{\beta}_{ln}(\boldsymbol{k})}{\left[\epsilon_m(\boldsymbol{k})-\epsilon_n(\boldsymbol{k})\right]\left[\epsilon_l(\boldsymbol{k})-\epsilon_n(\boldsymbol{k})\right]}-\frac{A^{\beta}_{mn}(\boldsymbol{k})A^{\alpha}_{nn}(\boldsymbol{k})}{\left[\epsilon_m(\boldsymbol{k})-\epsilon_n(\boldsymbol{k})\right]^2}+\frac{1}{\epsilon_m(\boldsymbol{k})-\epsilon_n(\boldsymbol{k})}i\partial_{k_{\alpha}}\left[\frac{A^{\beta}_{mn}(\boldsymbol{k})}{\epsilon_m(\boldsymbol{k})-\epsilon_n(\boldsymbol{k})}\right]\right\}+\text{h.c.}\\
    +&\left(\alpha\leftrightarrow\beta\right)
\end{split}$
}
\label{Eq:SecondResponseResult}
\end{equation}
with the right-hand side symmetrized with respect to the indices of $\alpha$ and $\beta$. The derivation is straightforward but tedious, see SM \cite{SupplementalM} for details. Integrand in the second order response [Eq.~(\ref{Eq:SecondResponseResult})] can be viewed as generalization of third rank quantum geometric tensors \cite{w761-8nf7,qscv-qxqt}.

To demonstrate the physical consequences of the general results above, we apply the response functions to the mean field bilayer EC system introduced at the beginning of the manuscript. For concreteness, we consider a simple nearest-neighbor tight-binding model on square lattice for each layer. Namely, we take $\xi(\boldsymbol{k})=-2t\left(\cos k_x +\cos k_y\right)$. In this particular case, the Bloch wavefunctions can be made purely real due to time-reversal symmetry, and the linear response function vanishes. We emphasize that the linear response function should generally be nonzero in low-symmetry systems without time reversal and rotation symmetries. 

The second order response function are nevertheless non-vanishing for bilayer EC. Direct calculation shows that
\begin{equation}
\begin{split}
    \chi_{z;\alpha\beta}^{(2)}=-\frac{1}{8}e^2\int\frac{d^2k}{(2\pi)^2}\frac{\Delta^2}{\epsilon^5(\boldsymbol{k})}\left[\partial^{2}_{k_{\alpha}k_{\beta}}\xi(\boldsymbol{k})-\frac{5}{2}\frac{\xi(\boldsymbol{k})-\mu}{\epsilon^{2}(\boldsymbol{k})}\partial_{k_{\alpha}}\xi(\boldsymbol{k})\partial_{k_{\beta}}\xi(\boldsymbol{k})\right]
    \label{Eq:Chi2_SquareLattice}
\end{split}
\end{equation}
where $\epsilon(\boldsymbol{k})=\sqrt{\left[\xi(\boldsymbol{k})-\mu\right]^2+\Delta^2}$. This integration can be carried out numerically. The result of $\chi^{(2)}_{z,xx}$ is shown in Fig.~\ref{fig:EC_chi}. Notice that $\chi^{(2)}_{z,\alpha\beta}=\chi^{(2)}_{z,xx}\delta_{\alpha\beta}$ in the current model. Simple analytical expression can be derived by making further approximations as below.
\end{widetext}

Let us make a few further approximations to explicitly reveal the properties of this second order response. Assuming the long wavelength limit $\xi(\boldsymbol{k})=-2t\cos k_x-2t\cos k_y\approx t\boldsymbol{k}^2-4t$ and a small order parameter $\Delta\ll|\mu+4t|\ll t$, the integration in Eq.~(\ref{Eq:Chi2_SquareLattice}) can be carried out (see SM \cite{SupplementalM}):
\begin{equation}
    \chi^{(2)}_{z,\alpha\beta}\approx C\times\frac{e^2}{\Delta^2}\delta_{\alpha\beta}
\end{equation}
where the constant is found to be $C \approx -0.01326$. Note that the order parameter $\Delta$ shows up in the denominator, which reveals the \emph{non-perturbative} nature of this effect. Indeed, the system is drastically different for $\Delta=0$ and $\Delta\neq0$. The major contribution to $\chi^{(2)}_{z,\alpha\beta}$ arises from electron and hole states involved in forming the bilayer EC, namely the states with momentum near the Fermi surface. On the other hand, the chemical potential provides only small corrections to the result above, see the numerical calculation in Fig.~\ref{fig:EC_chi}.

We expect the inverse-square power law in the second order response function $\chi^{(2)}_{z,\alpha\beta}\sim \Delta^{-2}$ to hold generally for bilayer ECs. Indeed, the EC induced quantum geometry is most drastic in the regime around quasi-particle gap. Meanwhile, the general result of second order response in Eq.~(\ref{Eq:SecondResponseResult}) should scale as inverse square with certain energy scale $[E]^{-2}$ from dimensional analysis. For states near the ``Fermi surface'', this energy scale is essentially given by the quasi-particle gap or the mean field order parameter. 

Consider an AC electric field $\boldsymbol{E}(t)=\hat{e}_xE_x\cos\omega t $, just to be concrete. The second order response function contains a DC component $\chi^{(2),\text{dc}}_{z,xx}$ and a second harmonic component $\chi^{(2),2\omega}_{z,xx}$, which are equal $\chi^{(2),\text{dc}}_{z,xx}=\chi^{(2),2\omega}_{z,\alpha\beta}=\frac{1}{2}\chi^{(2)}_{z,xx}$. The second harmonic component $\chi^{(2),2\omega}_{z,xx}$ dictates a vertical dipole oscillation with doubled frequency $2\omega$. Hence, our prediction is readily experimentally verifiable. Upon lowering temperature, we expect to observe a sharp increase in this second harmonic response across the critical temperature $T_c$. Then, at the temperature regime slightly below  $T_c$, the response function should exhibit a power law decrease as $\chi^{(2)}_{z,xx}\sim\frac{1}{T_c-T}$, given that the order parameter possesses a scaling $\Delta\sim\sqrt{T_c-T}$ as predicted by the Ginzburg-Landau theory.

\emph{Conclusion and future outlooks---}To conclude, we demonstrated that inter-layer excitonic coherence can generate a genuinely interaction-driven (``hidden'') quantum geometry even with trivial electron and hole bands. Using a quantum-kinetic framework, we showed that the hidden quantum geometry is directly encoded in the out-of-plane polarization's response to an in-plane AC electric field, and yields a characteristic quadratic (DC and second-harmonic) dipole signal with a distinctive scaling $\chi^{(2)}\propto\Delta^{-2}$ set by the excitonic gap.

We emphasize that our theory extends beyond bilayer exciton condensates and applies broadly to van der Waals stacked materials \cite{Geim2013vanderWaals,doi:10.1126/science.aac9439}, where one simply replaces the energies and wavefunctions in Eqs.~(\ref{Eq:LinearResponse})-(\ref{Eq:SecondResponseResult}) by those in the corresponding materials.

One interesting prediction of our theory is that an in-plane electric field can control the out-of-plane polarization through the responses in Eqs.~(\ref{Eq:LinearResponse})-(\ref{Eq:SecondResponseResult}). This suggests a route to electrically manipulate \emph{out-of-plane} spontaneous polarization via \emph{in-plane} electric fields, which finds potential applications in atomically thin ferroelectrics \cite{doi:10.1021/acsnano.7b02756,doi:10.1126/science.abd3230,doi:10.1126/science.abe8177,doi:10.1073/pnas.2115703118,PhysRevLett.130.146801} as a \emph{non-destructive} writing scheme for ferroelectric memories. Phenomenologically, the out-of-plane polarization $P_z$ of such van der Waals layered systems is determined by the Ginzburg-Landau equation:
\begin{equation}
    \alpha_2P_z+\alpha_4P_z^3=E_z^{\text{ext.}}+E_z(\delta P_z(\boldsymbol{E}))
    \label{Eq:OPP_control}
\end{equation}
where $\alpha_{2,4}$ are coefficients to be determined from microscopics. While an out-of-plane polarization is usually switched by a vertical electric field $E_z^{\text{ext}}$ in $z$-direction, our results in Eqs.~(\ref{Eq:LinearResponse})-(\ref{Eq:SecondResponseResult}) suggest an \emph{in-plane} field $\boldsymbol{E}$ can also serve this purpose: the in-plane field $\boldsymbol{E}$ drives a change in the out-of-plane polarization $\delta P_z(\boldsymbol{E})\sim\chi^{(2),\text{dc}}_{z;\alpha\beta}E_{\alpha}E_{\beta}$, which redistributes charges among different layers and generates an electric field $E_z(\delta P_z(\boldsymbol{E}))\sim\delta P_z(\boldsymbol{E})/\epsilon_{\text{dielectric}}$ with $\epsilon_{\text{dielectric}}$ being the dielectric constant. The thus generated electric field $E_z(\delta P_z(\boldsymbol{E}))$ then switches the spontaneous polarization as dictated by Eq.~(\ref{Eq:OPP_control}). 

\emph{Acknowledgements---}X.Y. acknowledges the support of Hong Kong Research Grants Council through Grant No. PDFS2425-6S02. B.T.Z. acknowledges the support of NSFC-Young Scientists Fund (No.12504194), Guangdong Provincial Quantum Science Strategic Initiative (No. GDZX2501004), Guangdong Provincial Talents Program (No. 2025D03J0006) and Start-up Fund of HKUST(GZ) (No. G0104000263).

\bibliography{main_ref}

\end{document}


\title{Supplemental Material: Hidden Berry Curvature Effects in Bilayer Exciton Condensate}

\author{Xuzhe Ying}
\affiliation{Department of Physics, Hong Kong University of Science and Technology, Clear Water Bay, Hong Kong, China}
\affiliation{Center for Theoretical Condensed Matter Physics, The Hong Kong University of Science and Technology, Clear Water Bay, Hong Kong, China}
\author{Benjamin T. Zhou} \thanks{Corresponding author: tongz@hkust-gz.edu.cn}
\affiliation{Quantum Science and Technology Center \& Thrust of Advanced Materials,
The Hong Kong University of Science and Technology (Guangzhou), 1 Duxue Rd., Nansha, Guangzhou, China}

\begin{abstract}
In this Supplemental Material, we provide detailed calculations for (i) Moyal diagonalization, (ii) linear response function, (iii) second order response function, (iv) application to simple model on square lattice.
\end{abstract}

\maketitle

\section{Moyal diagonalization}

Consider a tight-binding system under uniform external bias:
\begin{equation}
    H(\boldsymbol{r},\boldsymbol{r}^{\prime};t)=H_0(\boldsymbol{r}-\boldsymbol{r}^{\prime})+e\boldsymbol{E}\cdot\boldsymbol{r}\delta(\boldsymbol{r}-\boldsymbol{r}^{\prime})
\end{equation}
We perform Wigner transformation to rewrite this Hamiltonian in phase space:
\begin{equation}
\begin{split}
    H(\boldsymbol{k},\boldsymbol{r};t)=&\sum_{\boldsymbol{r}-\boldsymbol{r}^{\prime}}e^{i\boldsymbol{k}\cdot(\boldsymbol{r}-\boldsymbol{r}^{\prime})}H(\boldsymbol{r},\boldsymbol{r}^{\prime};t)
    =H_0(\boldsymbol{k})+e\boldsymbol{E}(t)\cdot\boldsymbol{r}
\end{split}
\end{equation}
We need to Moyal diagonalize this phase-space Hamiltonian:
\begin{equation}
    H(\boldsymbol{k},\boldsymbol{r};t)\star\mathbb{U}(\boldsymbol{k},\boldsymbol{r};t)=\mathbb{U}(\boldsymbol{k},\boldsymbol{r};t)\star\tilde{h}(\boldsymbol{k},\boldsymbol{r};t)
\end{equation}
where $\tilde{h}(\boldsymbol{k},\boldsymbol{r};t)$ is a diagonal matrix; the Moyal star operation is defined as:
\begin{equation}
    \star=\text{Exp}\left[-\frac{i\hbar}{2}\left(\overset{\leftarrow}{\partial_{\boldsymbol{r}}}\cdot\overset{\rightarrow}{\partial_{\boldsymbol{k}}}-\overset{\leftarrow}{\partial_{\boldsymbol{k}}}\cdot\overset{\rightarrow}{\partial_{\boldsymbol{r}}}\right)\right]
\end{equation}
In the maintext, we set $\hbar=1$, which is equivalent to saying that the value of $\hbar$ (including its unit) is absorb into the definition of momentum $\boldsymbol{k}$. Here, we need the explicit appearance of the Planck constant $\hbar$ as a perturbative parameter. For notational convenience, let us denote:
\begin{equation}
    \boldsymbol{\nabla}_{\boldsymbol{r}\boldsymbol{k}}:=\overset{\leftarrow}{\partial_{\boldsymbol{r}}}\cdot\overset{\rightarrow}{\partial_{\boldsymbol{k}}}-\overset{\leftarrow}{\partial_{\boldsymbol{k}}}\cdot\overset{\rightarrow}{\partial_{\boldsymbol{r}}}
\end{equation}

At zeroth order in $\hbar$, the equation reads:
\begin{equation}
    \left[H_0(\boldsymbol{k})+e\boldsymbol{E}(t)\cdot\boldsymbol{r}\right]\mathbb{U}^{(0)}(\boldsymbol{k},\boldsymbol{r};t)=\mathbb{U}^{(0)}(\boldsymbol{k},\boldsymbol{r};t)\tilde{h}^{(0)}(\boldsymbol{k},\boldsymbol{r};t)
\end{equation}
The solution is easy to find:
\begin{equation}
    \left\{\begin{split}
        &\mathbb{U}^{(0)}_n(\boldsymbol{k},\boldsymbol{r};t)=\mathbb{U}^{(0)}_n(\boldsymbol{k})=u_n(\boldsymbol{k})\\
        &\tilde{h}^{(0)}_n(\boldsymbol{k},\boldsymbol{r};t)=\epsilon_n(\boldsymbol{k})+e\boldsymbol{E}(t)\cdot\boldsymbol{r}
    \end{split}\right.
\end{equation}
Namely, the $n$-th eigenvector $\mathbb{U}^{(0)}_n(\boldsymbol{k},\boldsymbol{r};t)$ is given by the Bloch wavefunction of $n$-th quasi-particle band $u_n(\boldsymbol{k})$ and the corresponding eigenvalue is given by quasi-particle's kinetic energy plus the electric potential energy.

At first order in $\hbar$, the equation reads:
\begin{equation}
    \begin{split}
        &\left[H_0(\boldsymbol{k})+e\boldsymbol{E}(t)\cdot\boldsymbol{r}\right]\mathbb{U}^{(1)}(\boldsymbol{k},\boldsymbol{r};t)-\frac{i\hbar}{2}\left[H_0(\boldsymbol{k})+e\boldsymbol{E}(t)\cdot\boldsymbol{r}\right]\boldsymbol{\nabla}_{\boldsymbol{r}\boldsymbol{k}}\mathbb{U}^{(0)}(\boldsymbol{k})\\
        =&\mathbb{U}^{(1)}(\boldsymbol{k},\boldsymbol{r};t)\tilde{h}^{(0)}(\boldsymbol{k},\boldsymbol{r};t)+\mathbb{U}^{(0)}(\boldsymbol{k})\tilde{h}^{(1)}(\boldsymbol{k},\boldsymbol{r};t)-\frac{i\hbar}{2}\mathbb{U}^{(0)}(\boldsymbol{k})\boldsymbol{\nabla}_{\boldsymbol{r}\boldsymbol{k}}\tilde{h}^{(0)}(\boldsymbol{k},\boldsymbol{r};t)
    \end{split}
\end{equation}
This equation can be further simplified by noticing that $H(\boldsymbol{k},\boldsymbol{r};t)$ and $\tilde{h}^{(0)}(\boldsymbol{k},\boldsymbol{r};t)$ are at most linear in position $\boldsymbol{r}$. The simplified equation for the $n$-th eigenstate now reads:
\begin{equation}
    \left[H_0(\boldsymbol{k})+e\boldsymbol{E}(t)\cdot\boldsymbol{r}\right]\mathbb{U}^{(1)}(\boldsymbol{k},\boldsymbol{r};t)-\mathbb{U}^{(1)}(\boldsymbol{k},\boldsymbol{r};t)\left[\epsilon_n(\boldsymbol{k})+e\boldsymbol{E}(t)\cdot\boldsymbol{r}\right]=i\hbar e\boldsymbol{E}(t)\cdot\partial_{\boldsymbol{k}}u_n(\boldsymbol{k})
\end{equation}
The solution is given by:
\begin{equation}
    \left\{\begin{split}
        &\mathbb{U}^{(1)}_n(\boldsymbol{k};t)=\sum_{m\neq n}i\hbar eE_{\alpha}(t)u_m(\boldsymbol{k})\frac{u^{\dagger}_m(\boldsymbol{k})\partial_{k_{\alpha}}u_n(\boldsymbol{k})}{\epsilon_m(\boldsymbol{k})-\epsilon_n(\boldsymbol{k})}\\
        &\tilde{h}^{(1)}_n(\boldsymbol{k};t)=-i\hbar eE_{\alpha}u_n^{\dagger}(\boldsymbol{k})\partial_{k_{\alpha}}u_n(\boldsymbol{k})
    \end{split}
    \right.
\end{equation}

At second order in $\hbar$, the equation is given by:
\begin{equation}
    \begin{split}
        &\left[H_0(\boldsymbol{k})+e\boldsymbol{E}(t)\cdot\boldsymbol{r}\right]\frac{1}{2!}\left(-\frac{i\hbar}{2}\right)^2\boldsymbol{\nabla}_{\boldsymbol{r}\boldsymbol{k}}^2\mathbb{U}^{(0)}_n(\boldsymbol{k})+\left[H_0(\boldsymbol{k})+e\boldsymbol{E}(t)\cdot\boldsymbol{r}\right]\left(-\frac{i\hbar}{2}\right)^2\boldsymbol{\nabla}_{\boldsymbol{r}\boldsymbol{k}}\mathbb{U}^{(1)}_n(\boldsymbol{k};t)+\left[H_0(\boldsymbol{k})+e\boldsymbol{E}(t)\cdot\boldsymbol{r}\right]\mathbb{U}^{(2)}_n(\boldsymbol{k},\boldsymbol{r};t)\\
        =&\mathbb{U}^{(0)}_n(\boldsymbol{k})\frac{1}{2!}\left(-\frac{i\hbar}{2}\right)^2\boldsymbol{\nabla}_{\boldsymbol{r}\boldsymbol{k}}^2\left[\epsilon_n(\boldsymbol{k})+e\boldsymbol{E}(t)\cdot\boldsymbol{r}\right]+\mathbb{U}^{(0)}_n(\boldsymbol{k})\left(-\frac{i\hbar}{2}\right)\boldsymbol{\nabla}_{\boldsymbol{r}\boldsymbol{k}}\tilde{h}^{(1)}_n(\boldsymbol{k};t)+\mathbb{U}^{(0)}_n(\boldsymbol{k})\tilde{h}^{(2)}_n(\boldsymbol{k},\boldsymbol{r};t)\\
        +&\mathbb{U}^{(1)}_n(\boldsymbol{k})\left(-\frac{i\hbar}{2}\right)\boldsymbol{\nabla}_{\boldsymbol{r}\boldsymbol{k}}\left[\epsilon_n(\boldsymbol{k})+e\boldsymbol{E}(t)\cdot\boldsymbol{r}\right]+\mathbb{U}^{(1)}_n(\boldsymbol{k};t)\tilde{h}^{(1)}_n(\boldsymbol{k};t)+\mathbb{U}^{(2)}_n(\boldsymbol{k};t)\tilde{h}^{(0)}_n(\boldsymbol{k};t)
    \end{split}
\end{equation}
This equation can be further simplified:
\begin{equation}
    \begin{split}
        &\left[H_0(\boldsymbol{k})+e\boldsymbol{E}(t)\cdot\boldsymbol{r}\right]\mathbb{U}^{(2)}_n(\boldsymbol{k},\boldsymbol{r};t)-\mathbb{U}^{(2)}_n(\boldsymbol{k};t)\left[\epsilon_n(\boldsymbol{k})+e\boldsymbol{E}(t)\cdot\boldsymbol{r}\right]\\
        =&i\hbar e\boldsymbol{E}(t)\cdot\partial_{\boldsymbol{k}}\mathbb{U}^{(1)}_n(\boldsymbol{k};t)+\mathbb{U}^{(1)}_n(\boldsymbol{k};t)\tilde{h}^{(1)}_n(\boldsymbol{k};t)+\mathbb{U}^{(0)}_n(\boldsymbol{k})\tilde{h}^{(2)}_n(\boldsymbol{k},\boldsymbol{r};t)
    \end{split}
\end{equation}
The solution is given by:
\begin{equation}
    \left\{\begin{split}
        \mathbb{U}^{(2)}_n(\boldsymbol{k};t)=&-\frac{1}{2}\hbar^2e^2E_{\alpha}(t)E_{\beta}(t)u_n(\boldsymbol{k})\sum_{m\neq n}\frac{\partial_{k_{\alpha}}u_n^{\dagger}(\boldsymbol{k})\ u_m(\boldsymbol{k})u_m^{\dagger}(\boldsymbol{k})\ \partial_{k_{\beta}}u_n(\boldsymbol{k})}{\left[\epsilon_m(\boldsymbol{k})-\epsilon_n(\boldsymbol{k})\right]^2}\\
        &-\sum_{m\neq n}u_m(\boldsymbol{k})\frac{\hbar^2e^2E_{\alpha}(t)E_{\beta}(t)}{\epsilon_m(k)-\epsilon_n(k)}\left\{\sum_{l\neq n}u^{\dagger}_m(\boldsymbol{k})\partial_{k_{\alpha}}u_l(\boldsymbol{k})\frac{u_{l}^{\dagger}(\boldsymbol{k})\partial_{k_{\beta}}u_n(\boldsymbol{k})}{\epsilon_l(\boldsymbol{k})-\epsilon_n(\boldsymbol{k})}+\partial_{k_{\alpha}}\left[\frac{u^{\dagger}_m(\boldsymbol{k})\partial_{k_{\beta}}u_n(\boldsymbol{k})}{\epsilon_m(\boldsymbol{k})-\epsilon_n(\boldsymbol{k})}\right]\right.\\
        &\ \ \ \ \ \ \ \ \ \ \ \ \ \ \ \ \ \ \ \ \ \ \ \ \ \ \ \ \ \ \ \ \ \ \ \ \ \ \ \ \ \ \ \ \ \ \ \ \ \ \ \ \ \ \left.+ \frac{\partial_{k_{\beta}}u_m^{\dagger}(\boldsymbol{k})\ u_n(\boldsymbol{k})u_n^{\dagger}(\boldsymbol{k})\ \partial_{k_{\alpha}}u_n(\boldsymbol{k})}{\epsilon_m(\boldsymbol{k})-\epsilon_n(\boldsymbol{k})}\right\}\\
        \tilde{h}^{(2)}_n(\boldsymbol{k};t)=&-i\hbar eE_{\alpha}(t)u_n^{\dagger}(\boldsymbol{k})\partial_{k_{\alpha}}\mathbb{U}^{(1)}(\boldsymbol{k};t)
    \end{split}
    \right.
\end{equation}

\section{Linear response of an exciton insulator}

\subsection{Result from kinetic equations}

For an exciton insulator, the distribution function remains the equilibrium Fermi distribution. The response comes from the change in the wavefunction, and hence the observables.

It is easy to write down the expectation value of out-of-plane polarization to linear order in electric field:
\begin{equation}
    P_z^{(1)}(t)=\sum_n\int\frac{d^2k}{(2\pi)^2}f_n^{\text{eq}}(\boldsymbol{k})\times\left[\mathbb{U}^{(1),\dagger}_n(\boldsymbol{k};t)\ \hat{z}\ \mathbb{U}^{(0)}_n(\boldsymbol{k};t)+\mathbb{U}^{(0),\dagger}_n(\boldsymbol{k};t)\ \hat{z}\ \mathbb{U}^{(1)}_n(\boldsymbol{k};t)\right]
\end{equation}
Substituting in the expression for $\mathbb{U}^{(1)}_n(\boldsymbol{k};t)$, we obtain the linear response function:
\begin{equation}
    P_{z}^{(1)}(t)=\hbar e E_{\alpha}(t)\sum_{n}\sum_{m\neq n}\int\frac{d^2k}{(2\pi)^2}\frac{\hat{z}_{nm}(\boldsymbol{k})A^{\alpha}_{mn}(\boldsymbol{k})+A^{\alpha}_{nm}(\boldsymbol{k})\hat{z}_{mn}(\boldsymbol{k})}{\epsilon_m(\boldsymbol{k})-\epsilon_n(\boldsymbol{k})}f_n(\boldsymbol{k})
\end{equation}
We compare this result with the Kubo formula as below.

\subsection{Kubo formula result}

The Kubo formula calculation amounts to computing the correlation function:
\begin{equation}
    \Pi_{\hat{z}\hat{j}_{\alpha}}(i\Omega,\boldsymbol{q}=0)=\text{Tr}\ \left[G_0(i\omega+i\Omega,\boldsymbol{k})\ \hat{z}\ G_0(i\omega,\boldsymbol{k})\ \hat{j}_{\alpha}\right]
\end{equation}
Utilizing Lehmann representation, this polarization can be written as:
\begin{equation}
    \Pi_{\hat{z}\hat{j}_{\alpha}}(i\Omega,\boldsymbol{q}=0)=\sum_{i\omega}\sum_{m,n}\int\frac{d^2k}{(2\pi)^2}\frac{u^{\dagger}_m(\boldsymbol{k})\hat{z}u_n(\boldsymbol{k})u^{\dagger}_n(\boldsymbol{k})\hat{j}_{\alpha}u_m(\boldsymbol{k})}{\left[i\omega+i\Omega-\epsilon_m(\boldsymbol{k})\right]\left[i\omega-\epsilon_n(\boldsymbol{k})\right]}
\end{equation}
After summation over Matsubara frequence $i\omega$, the result is:
\begin{equation}
    \Pi_{\hat{z}\hat{j}_{\alpha}}(i\Omega,\boldsymbol{q}=0)=\sum_{m,n}\int\frac{d^2k}{(2\pi)^2}u^{\dagger}_m(\boldsymbol{k})\hat{z}u_n(\boldsymbol{k})u^{\dagger}_n(\boldsymbol{k})\hat{j}_{\alpha}u_m(\boldsymbol{k})\frac{f_m^{\text{eq}}(\boldsymbol{k})-f_n^{\text{eq}}(\boldsymbol{k})}{-i\Omega+\epsilon_m(\boldsymbol{k})-\epsilon_{n}(\boldsymbol{k})}
\end{equation}
The next step is to perform the analytical continuation:
\begin{equation}
    \Pi_{\hat{z}\hat{j}_{\alpha}}(i\Omega\rightarrow\omega+i0^{+},\boldsymbol{q}=0)=\sum_{m,n}\int\frac{d^2k}{(2\pi)^2}u^{\dagger}_m(\boldsymbol{k})\hat{z}u_n(\boldsymbol{k})u^{\dagger}_n(\boldsymbol{k})\hat{j}_{\alpha}u_m(\boldsymbol{k})\frac{f_m^{\text{eq}}(\boldsymbol{k})-f_n^{\text{eq}}(\boldsymbol{k})}{-\omega-i0^{+}+\epsilon_m(\boldsymbol{k})-\epsilon_{n}(\boldsymbol{k})}
\end{equation}
Lastly, we can compute the response function as:
\begin{equation}
    \chi_{\hat{z}\hat{j}_{\alpha}}^{(1)}(\omega)=\frac{1}{i\omega}\left[\Pi_{\hat{z}\hat{j}_{\alpha}}(\omega+i0^+,\boldsymbol{q}=0)-\Pi_{\hat{z}\hat{j}_{\alpha}}(\omega=0,\boldsymbol{q}=0)\right]
\end{equation}
In the low-frequency limit, the result above is:
\begin{equation}
    \chi_{\hat{z}\hat{j}_{\alpha}}^{(1)}(\omega)=-i\sum_{m,n}\int\frac{d^2k}{(2\pi)^2}u^{\dagger}_m(\boldsymbol{k})\hat{z}u_n(\boldsymbol{k})u^{\dagger}_n(\boldsymbol{k})\hat{j}_{\alpha}u_m(\boldsymbol{k})\frac{f_m^{\text{eq}}(\boldsymbol{k})-f_n^{\text{eq}}(\boldsymbol{k})}{\left[\epsilon_m(\boldsymbol{k})-\epsilon_{n}(\boldsymbol{k})\right]^2}
\end{equation}
This expression can be further simplified:
\begin{equation}
    \chi_{\hat{z}\hat{j}_{\alpha}}^{(1)}(\omega)=-i\sum_{n}\int\frac{d^2k}{(2\pi)^2}\sum_{m\neq n}\frac{u^{\dagger}_n(\boldsymbol{k})\hat{z}u_m(\boldsymbol{k})u^{\dagger}_m(\boldsymbol{k})\hat{j}_{\alpha}u_n(\boldsymbol{k})-u^{\dagger}_m(\boldsymbol{k})\hat{z}u_n(\boldsymbol{k})u^{\dagger}_n(\boldsymbol{k})\hat{j}_{\alpha}u_m(\boldsymbol{k})}{\left[\epsilon_m(\boldsymbol{k})-\epsilon_{n}(\boldsymbol{k})\right]^2}f_n^{\text{eq}}(\boldsymbol{k})
\end{equation}
Let us specify the current operator $\hat{j}_{\alpha}=-e\partial_{k_{\alpha}}H_0(\boldsymbol{k})$ and utilize the following relation:
\begin{equation}
    u^{\dagger}_n(\boldsymbol{k})\partial_{k_{\alpha}}H_0(\boldsymbol{k})u_m(\boldsymbol{k})=-\left[\epsilon_n(\boldsymbol{k})-\epsilon_m(\boldsymbol{k})\right]u^{\dagger}_n(\boldsymbol{k})\partial_{k_{\alpha}}u_m(\boldsymbol{k})
\end{equation}
We can reproduce the result from the kinetic equation for linear response.

\section{Second order response function}

The second order result is formally given by:
\begin{equation}
    P_z^{(2)}(t)=\sum_n\int\frac{d^2k}{(2\pi)^2}\sum_{i=0,1,2}\mathbb{U}^{(i),\dagger}_n(\boldsymbol{k};t)\ \hat{z}\ \mathbb{U}^{(2-i)}_n(\boldsymbol{k};t)\ f_n^{\text{eq}(\boldsymbol{k})}
\end{equation}

The ``$(1,1)$'' component is given by:
\begin{equation}
    \begin{split}
        P_z^{(2),(1,1)}(t)=&\sum_n\int\frac{d^2k}{(2\pi)^2}\mathbb{U}^{(1),\dagger}_n(\boldsymbol{k};t)\ \hat{z}\ \mathbb{U}^{(1)}_n(\boldsymbol{k};t)\ f_n^{\text{eq}(\boldsymbol{k})}\\
        =&\hbar^2e^2E_{\alpha}(t)E_{\beta}(t)\sum_{n}\sum_{m_1,m_2\neq n}\int\frac{d^2k}{(2\pi)^2}\frac{A^{\alpha}_{nm_1}(\boldsymbol{k})\hat{z}_{m_1m_2}(\boldsymbol{k})A^{\beta}_{m_2n}(\boldsymbol{k})}{\left[\epsilon_{m_1}(\boldsymbol{k})-\epsilon_{n}(\boldsymbol{k})\right]\left[\epsilon_{m_2}(\boldsymbol{k})-\epsilon_{n}(\boldsymbol{k})\right]}f_n^{\text{eq}}(\boldsymbol{k})
    \end{split}
\end{equation}

The ``$(2,0)+(0,2)$'' component is given by:
\begin{equation}
    \begin{split}
        P_z^{(2),(2,0)+(0,2)}(t)=&\sum_n\int\frac{d^2k}{(2\pi)^2}\left[\mathbb{U}^{(2),\dagger}_n(\boldsymbol{k};t)\ \hat{z}\ \mathbb{U}^{(0)}_n(\boldsymbol{k};t)+\mathbb{U}^{(0),\dagger}_n(\boldsymbol{k};t)\ \hat{z}\ \mathbb{U}^{(2)}_n(\boldsymbol{k};t)\right]\ f_n^{\text{eq}(\boldsymbol{k})}\\
        =&-\frac{1}{2}\hbar^2e^2E_{\alpha}(t)E_{\beta}(t)\sum_n\int\frac{d^2k}{(2\pi)^2}f_n^{\text{eq}}(\boldsymbol{k})\hat{z}_{nn}(\boldsymbol{k})\sum_{m\neq n}\frac{A^{\alpha}_{nm}(\boldsymbol{k})A^{\beta}_{mn}(\boldsymbol{k})+A^{\beta}_{nm}(\boldsymbol{k})A^{\alpha}_{mn}(\boldsymbol{k})}{\left[\epsilon_{m}(\boldsymbol{k})-\epsilon_n(\boldsymbol{k})\right]^2}\\
        +&\hbar^2e^2E_{\alpha}(t)E_{\beta}(t)\sum_{n}\int\frac{d^2k}{(2\pi)^2}f_n^{\text{eq}}(\boldsymbol{k})\sum_{m\neq n}\hat{z}_{nm}\sum_{l\neq n}\frac{A^{\alpha}_{ml}(\boldsymbol{k})A^{\beta}_{ln}(\boldsymbol{k})}{\left[\epsilon_m(\boldsymbol{k})-\epsilon_n(\boldsymbol{k})\right]\left[\epsilon_l(\boldsymbol{k})-\epsilon_n(\boldsymbol{k})\right]}+\text{h.c.}\\
        -&\hbar^2e^2E_{\alpha}(t)E_{\beta}(t)\sum_{n}\int\frac{d^2k}{(2\pi)^2}f_n^{\text{eq}}(\boldsymbol{k})\sum_{m\neq n}\hat{z}_{nm}\frac{A^{\beta}_{mn}(\boldsymbol{k})A^{\alpha}_{nn}(\boldsymbol{k})}{\left[\epsilon_m(\boldsymbol{k})-\epsilon_n(\boldsymbol{k})\right]^2}+\text{h.c.}\\
        +&\hbar^2e^2E_{\alpha}(t)E_{\beta}(t)\sum_{n}\int\frac{d^2k}{(2\pi)^2}f_n^{\text{eq}}(\boldsymbol{k})\sum_{m\neq n}\frac{\hat{z}_{nm}}{\epsilon_m(\boldsymbol{k})-\epsilon_n(\boldsymbol{k})}i\partial_{k_{\alpha}}\left[\frac{A^{\beta}_{mn}(\boldsymbol{k})}{\epsilon_m(\boldsymbol{k})-\epsilon_n(\boldsymbol{k})}\right]+\text{h.c.}
    \end{split}
\end{equation}
Hence, the total second order response function is given by
\begin{equation}
\begin{split}
    \chi_{z;\alpha\beta}^{(2)}/\hbar^2e^2=&\frac{1}{2}\sum_{n}\int\frac{d^2k}{(2\pi)^2}f_n^{\text{eq}}(\boldsymbol{k})\sum_{m_1,m_2\neq n}\frac{A^{\alpha}_{nm_1}(\boldsymbol{k})\hat{z}_{m_1m_2}(\boldsymbol{k})A^{\beta}_{m_2n}(\boldsymbol{k})+\left(\alpha\leftrightarrow\beta\right)}{\left[\epsilon_{m_1}(\boldsymbol{k})-\epsilon_{n}(\boldsymbol{k})\right]\left[\epsilon_{m_2}(\boldsymbol{k})-\epsilon_{n}(\boldsymbol{k})\right]}\\
    -&\frac{1}{2}\sum_n\int\frac{d^2k}{(2\pi)^2}f_n^{\text{eq}}(\boldsymbol{k})\hat{z}_{nn}(\boldsymbol{k})\sum_{m\neq n}\frac{A^{\alpha}_{nm}(\boldsymbol{k})A^{\beta}_{mn}(\boldsymbol{k})+A^{\beta}_{nm}(\boldsymbol{k})A^{\alpha}_{mn}(\boldsymbol{k})}{\left[\epsilon_{m}(\boldsymbol{k})-\epsilon_n(\boldsymbol{k})\right]^2}\\
    +&\frac{1}{2}\sum_{n}\int\frac{d^2k}{(2\pi)^2}f_n^{\text{eq}}(\boldsymbol{k})\sum_{m\neq n}\hat{z}_{nm}\left\{\sum_{l\neq n}\frac{A^{\alpha}_{ml}(\boldsymbol{k})A^{\beta}_{ln}(\boldsymbol{k})}{\left[\epsilon_m(\boldsymbol{k})-\epsilon_n(\boldsymbol{k})\right]\left[\epsilon_l(\boldsymbol{k})-\epsilon_n(\boldsymbol{k})\right]}+\left(\alpha\leftrightarrow\beta\right)\right\}+\text{h.c.}\\
    -&\frac{1}{2}\sum_{n}\int\frac{d^2k}{(2\pi)^2}f_n^{\text{eq}}(\boldsymbol{k})\sum_{m\neq n}\hat{z}_{nm}\left\{\frac{A^{\beta}_{mn}(\boldsymbol{k})A^{\alpha}_{nn}(\boldsymbol{k})}{\left[\epsilon_m(\boldsymbol{k})-\epsilon_n(\boldsymbol{k})\right]^2}+\left(\alpha\leftrightarrow\beta\right)\right\}+\text{h.c.}\\
    +&\frac{1}{2}\sum_{n}\int\frac{d^2k}{(2\pi)^2}f_n^{\text{eq}}(\boldsymbol{k})\sum_{m\neq n}\left\{\frac{\hat{z}_{nm}}{\epsilon_m(\boldsymbol{k})-\epsilon_n(\boldsymbol{k})}i\partial_{k_{\alpha}}\left[\frac{A^{\beta}_{mn}(\boldsymbol{k})}{\epsilon_m(\boldsymbol{k})-\epsilon_n(\boldsymbol{k})}\right]+\left(\alpha\leftrightarrow\beta\right)\right\}+\text{h.c.}
\end{split}
\end{equation}

\section{Application to a simple model on square lattice}

We apply the theory above to a simple model defined on square lattice. We start from the following mean field Hamiltonian:
\begin{equation}
    H_0(\boldsymbol{k})=\begin{bmatrix}
        -\xi(\boldsymbol{k})+\mu & \Delta\\
        \Delta & \xi(\boldsymbol{k})-\mu
    \end{bmatrix}
\end{equation}
This mean field Hamiltonian contains two quasi-particle bands, whose spectra and wavefunctions are given below:
\begin{equation}
	\left\{\begin{split}
		& \epsilon_+(\boldsymbol{k})=+\sqrt{\left[\xi(\boldsymbol{k})-\delta\mu\right]^2+\Delta^2},\ \ \ u_+(\boldsymbol{k})=\mathcal{N}_{\boldsymbol{k}}\begin{bmatrix}
				\xi(\boldsymbol{k})-\mu+\epsilon(\boldsymbol{k})\\
				\Delta
		\end{bmatrix},\\
		& \epsilon_-(\boldsymbol{k})=-\sqrt{\left[\xi(\boldsymbol{k})-\delta\mu\right]^2+\Delta^2},\ \ \ u_-(\boldsymbol{k})=\mathcal{N}_{\boldsymbol{k}}\begin{bmatrix}
				-\Delta\\
				\xi(\boldsymbol{k})-\mu+\epsilon(\boldsymbol{k})
		\end{bmatrix}
	\end{split}\right.
\end{equation}
with
\begin{equation}
    \xi(\boldsymbol{k})=-2t\cos k_x-2t\cos k_y,\ \ \epsilon(\boldsymbol{k})=\sqrt{\left[\xi(\boldsymbol{k})-\delta\mu\right]^2+\Delta^2},\ \ \mathcal{N}_{\boldsymbol{k}}=\left[2\epsilon(\boldsymbol{k})(\epsilon(\boldsymbol{k})+\xi(\boldsymbol{k})-\mu)\right]^{\frac{1}{2}}
\end{equation}
Then, we can compute the quantum geometric quantities straightforwardly (note $\hat{z}:=-\frac{1}{2}\sigma_z$):
\begin{equation}
\begin{split}
    &\hat{z}_{-+}(\boldsymbol{k})=\hat{z}_{+-}(\boldsymbol{k})=\frac{\Delta}{2\epsilon(\boldsymbol{k})},\ \ \ \hat{z}_{--}(\boldsymbol{k})=-\hat{z}_{++}(\boldsymbol{k})=\frac{\xi(\boldsymbol{k})-\mu}{2\epsilon(\boldsymbol{k})}\\
    &A^{\alpha}_{mn}(\boldsymbol{k})=\begin{bmatrix}
        u_-^{\dagger}(\boldsymbol{k})i\partial_{k_{\alpha}}u_-(\boldsymbol{k}) & u_-^{\dagger}(\boldsymbol{k})i\partial_{k_{\alpha}}u_+(\boldsymbol{k})\\
        u_+^{\dagger}(\boldsymbol{k})i\partial_{k_{\alpha}}u_-(\boldsymbol{k}) & u_+^{\dagger}(\boldsymbol{k})i\partial_{k_{\alpha}}u_+(\boldsymbol{k})
    \end{bmatrix}=\begin{bmatrix}
        0 & -\frac{i\Delta\partial_{k_{\alpha}}\xi(\boldsymbol{k})]}{2\epsilon^2(\boldsymbol{k})}\\
        \frac{i\Delta\partial_{k_{\alpha}}\xi(\boldsymbol{k})}{2\epsilon^2(\boldsymbol{k})} & 0
    \end{bmatrix}
\end{split}
\end{equation}
With the quantum geometric information, we can compute the response functions of an exciton insulator. It is easy to show that linear response function vanishes for the simple model under consideration. Then, we focus on the second order response.

The expression for the second order response can be simplified for the current case with two quasi-particle's band:
\begin{equation}
\begin{split}
    \chi^{(2)}_{z;\alpha\beta}/e^2=&\frac{1}{2}\int\frac{d^2k}{(2\pi)^2}\frac{\hat{z}_{++}(\boldsymbol{k})-\hat{z}_{--}(\boldsymbol{k})}{\left[\epsilon_+(\boldsymbol{k})-\epsilon_-(\boldsymbol{k})\right]^2}\left[A^{\alpha}_{-+}(\boldsymbol{k})A^{\beta}_{-+}(\boldsymbol{k})+A^{\beta}_{-+}(\boldsymbol{k})A^{\alpha}_{-+}(\boldsymbol{k})\right]\\
    +&\frac{1}{2}\int\frac{d^2k}{(2\pi)^2}\frac{\hat{z}_{-+}(\boldsymbol{k})}{\epsilon_+(\boldsymbol{k})-\epsilon_-(\boldsymbol{k})}\left[i\frac{\partial}{\partial k_{\alpha}}\frac{A^{\beta}_{+-}(\boldsymbol{k})}{\epsilon_+(\boldsymbol{k})-\epsilon_-(\boldsymbol{k})}+(\alpha\leftrightarrow\beta)\right]+\text{h.c.}
\end{split}
\end{equation}
Substituting in the expressions for the quantum geometric quantities, we find:
\begin{equation}
    \chi^{(2)}_{z;\alpha\beta}/e^2=-\frac{1}{8}\int\frac{d^2k}{(2\pi)^2}\frac{\Delta^2}{\epsilon^5(\boldsymbol{k})}\left[\partial^2_{k_{\alpha}k_{\beta}}\xi(\boldsymbol{k})-\frac{5}{2}\frac{\xi(\boldsymbol{k})-\mu}{\epsilon^2(\boldsymbol{k})}\partial_{k_{\alpha}}\xi(\boldsymbol{k})\partial_{k_{\beta}}\xi(\boldsymbol{k})\right]
\end{equation}
To reveal the details of such response function, let us make further approximations. First, we put the chemical potential $\mu$ to be near the band edge and take the long wavelength limit:
\begin{equation}
    \xi(\boldsymbol{k})-\mu\approx\frac{\boldsymbol{k}^2}{2m}-\delta\mu
\end{equation}
where the effective mass is $m^{-1}=2t$ and $\delta\mu=-4t-\mu$. For this to be a good approximation, we further assume a small order parameter $\Delta\ll|\delta\mu|\ll t$.

Then, it is straightforward to compute, for instance $\chi^{(2)}_{z;xx}$:
\begin{equation}
    \chi^{(2)}_{z;xx}=-\frac{e^2}{8}\int\frac{d^2k}{(2\pi)^2}\frac{\Delta^2}{\left[\left(\frac{\boldsymbol{k}^2}{2m}-\delta\mu\right)^2+\Delta^2\right]^{\frac{5}{2}}}\left[\frac{1}{m}-\frac{5}{2}\frac{\boldsymbol{k}^2/2m-\delta\mu}{\left(\boldsymbol{k}^2/2m-\delta\mu\right)^2+\Delta^2}\frac{k_x^2}{m^2}\right]
\end{equation}
We will evaluate this expression below. First step is to replace $k_x^2\rightarrow\boldsymbol{k^2}/2$ and factor out $\frac{1}{m}$:
\begin{equation}
    \chi^{(2)}_{z;xx}=-\frac{e^2}{8}\int\frac{d^2k}{(2\pi)^2}\frac{1}{m}\frac{\Delta^2}{\left[\left(\frac{\boldsymbol{k}^2}{2m}-\delta\mu\right)^2+\Delta^2\right]^{\frac{5}{2}}}\left[1-\frac{5}{2}\frac{\boldsymbol{k}^2/2m-\delta\mu}{\left(\boldsymbol{k}^2/2m-\delta\mu\right)^2+\Delta^2}\frac{\boldsymbol{k}^2}{2m}\right]
\end{equation}
Then, we replace the momentum integration by energy integration $\xi=\boldsymbol{k}^2/2m-\delta\mu$ and integrate over $\mathbb{R}$:
\begin{equation}
\begin{split}
    \chi^{(2)}_{z;xx}\approx&-\frac{e^2\Delta^2}{16\pi}\int_{-\infty}^{\infty}d\xi\frac{1}{\left(\xi^2+\Delta^2\right)^{\frac{5}{2}}}\left(1-\frac{5}{2}\frac{\xi}{\xi^2+\Delta^2}(\xi+\delta\mu)\right)\\
    =&-\frac{e^2\Delta^2}{16\pi}\int_{-\infty}^{\infty}d\xi\frac{1}{\left(\xi^2+\Delta^2\right)^{\frac{5}{2}}}\left(1-\frac{5}{2}\frac{\xi^2}{\xi^2+\Delta^2}\right)
\end{split}
\end{equation}
To perform this integration explicitly, we make a change of variable $\xi=\Delta\sinh x$:
\begin{equation}
\begin{split}
    \chi^{(2)}_{z;xx}\approx&-\frac{e^2}{16\pi\Delta^2}\int_{-\infty}^{\infty}dx\frac{1}{\cosh ^4x}\left(1-\frac{5\sinh^2x}{2\cosh^2x}\right)
    \approx-0.01326\frac{e^2}{\Delta^2}
\end{split}
\end{equation}
which is the result reported in the maintext.